\documentclass[a4paper,12pt,titlepage]{article}

\usepackage{graphics}
\usepackage{doi}
\usepackage[nooneline]{caption}
\usepackage{siunitx}
\usepackage{upgreek}
\usepackage{setspace}
\usepackage{hyperref}

\usepackage[symbol]{footmisc}

\newlength\figurewidth
\parindent=0pt{}
\parskip=\baselineskip

\begin{document}

\title{Derivation of an Improved Semi-empirical Expression for the Re-ionisation Background in Low Energy Ion Scattering Spectra}

\makeatletter
\let\anontitle\@title
\makeatother

\author{
\parbox{\textwidth}{\normalsize
H. R. Koslowski$^{\ast}$ and Ch. Linsmeier
\\
\\
\normalsize{}
Forschungszentrum J\"ulich GmbH, Institut f\"ur Energie- und Klimaforschung -- Plasmaphysik, 52425 J\"ulich, Germany\\
\\
$^{\ast}$ Email: h.r.koslowski@fz-juelich.de
}}

\date{\today}

\maketitle

\abstract{
Low energy ion scattering is a technique to detect the energy of ions which are scattered from a surface.
For noble gas ions, it is predominantly sensitive to the topmost surface layer due to strong neutralisation processes.
Depending on the combination of projectile ion and target material, the scattering spectra can exhibit contributions resulting from multiple scattering processes in deeper layers when probing ions are re-ionised on the exiting trajectory.
These events cause a pronounced continuum located toward lower scattering energies with respect to the direct scattering peak.
In a previous work a semi-empirical formula has been given which allows fitting and derivation of quantitative information from the measured spectra [Nelson 1986 J. Vac. Sci. Technol. A 4 1567-1569].
Based on the former work an improved formula is derived which has less numerical artefacts and is numerically more stable.
}

\vfill

\newpage
\section{Introduction}
Low energy ion scattering from surfaces is a well established method to study the elemental composition of the topmost surface layer.
The method has been described first from Smith in 1967 \cite{smi67}.
An overview on the fundamental physics, possible applications, and applied apparatuses and technologies for production, scattering, and detection of the ions can be found in the review paper by Niehus \cite{nie93} (and, of course, in many other reviews, too).
Charge exchange processes play an important role in ion-surface collisions.
Neutralisation and ionisation processes between noble gas ions (which are usually used as projectiles) and the surface during approach and exit have been studied in great detail and overviews can be found in recent review papers by Monreal \cite{mon14} or Brongersma \cite{bro07}.
The most prominent process in this context is Auger neutralisation, but (quasi) resonant neutralisation and collision induced neutralisation can happen, too.
The actual occurrence of any of these processes depends on the properties of the collision system, e.g. the work function of the solid, the energy levels and their corresponding shifts of the projectile ion, and the collision energy.

An important feature of the scattering process from e.g. metallic surfaces is that most of the ions are neutralised on the incoming trajectory and only a small fraction of the impinging particles remains ionised after the scattering event.
The large neutralisation probability during the scattering process, or with other words, the small probability that the projectile ion remains charged after the collision, is crucial for the large sensitivity to scattering events from the topmost surface layer, because the survival probability for ions scattered multiple times at deeper layers is strongly reduced.
However, for certain combinations of projectile ions and target species the scattering spectra show a very asymmetric direct scattering peak with a pronounced low energy tail.
A considerable number of scattered projectile ions is detected at smaller energy than the single scattering peak, suggesting that these particles underwent a higher energy loss due to multiple binary collisions on their path through the solid surface region.
The origin of this continuum is discussed in \cite{bro07} and more recently \cite{bru19,bru20} deal especially with the role of oxygen for the re-ionisation.
The most likely and consistent explanation of these findings is that a projectile ion is scattered multiple times at deeper layers below the surface and re-ionised when leaving the surface.
Well known examples of projectile/target combinations which exhibit a re-ionisation background are the scattering of He$^{+}$ ions from e.g. Al, Ta, or W surfaces.
More detailed list can be found in \cite{bro07,sou87,tho86}.

With increasing number of collisions the probing particles reach deeper layers and lose more energy.
This re-ionisation continuum below the scattering peak can even yield depth information \cite{bei01,pri11}.
The assumption that re-ionisation is the responsible mechanism for the low energy tails is further supported by the existence of an energy threshold which results from the minimal energy required for a sufficiently close approach of nuclei to allow electronic orbital curve-crossings to occur \cite{bro07,bau78}.

\section{Example of a re-ionisation continuum}
A typical low energy ion scattering spectrum which clearly shows the re-ionisation continuum is depicted in figure \ref{fig:spec}.
Here, He$^{+}$ ions with an energy of \SI{1}{keV} are scattered at a poly-crystalline W surface which is heated to a temperature of $T=\SI{860}{K}$, and detected at a scattering angle of \ang{140}.
The sample is kept at an elevated temperature in order to remove oxygen from the surface and give a clean spectrum.
The symmetric peak located at an energy of $E_{0}=\SI{926}{eV}$ at the right side of the spectrum results from single scattering and can generally be well fitted by a Gaussian profile.
The use of Lorentzian or mixed Gaussian/Lorentzian functions has been suggested in the work by Nelson \cite{nel86}, although this has been adopted from other techniques, mostly electron spectroscopy.
For the scattering of He$^{+}$ on Al it has been shown in \cite{run11} that the single scattering peak could be modelled as a superposition of two Gaussians which correspond to ions either remaining charged during the whole collision or being re-ionised in a close collision after Auger neutralisation on the incoming trajectory.
The long left wing tail is caused by re-ionisation events and has a rather unique shape.
It starts at the high energy side at the peak location and extends with a decreasing amplitude down to a lower energy value below which the re-ionisation processes have ceased.
Note that the energy scale starts at 400 eV, below which almost no ion intensity is detected.

\section{Derivation of the background expression}
\begin{figure}
\resizebox{\figurewidth}{!}{\includegraphics{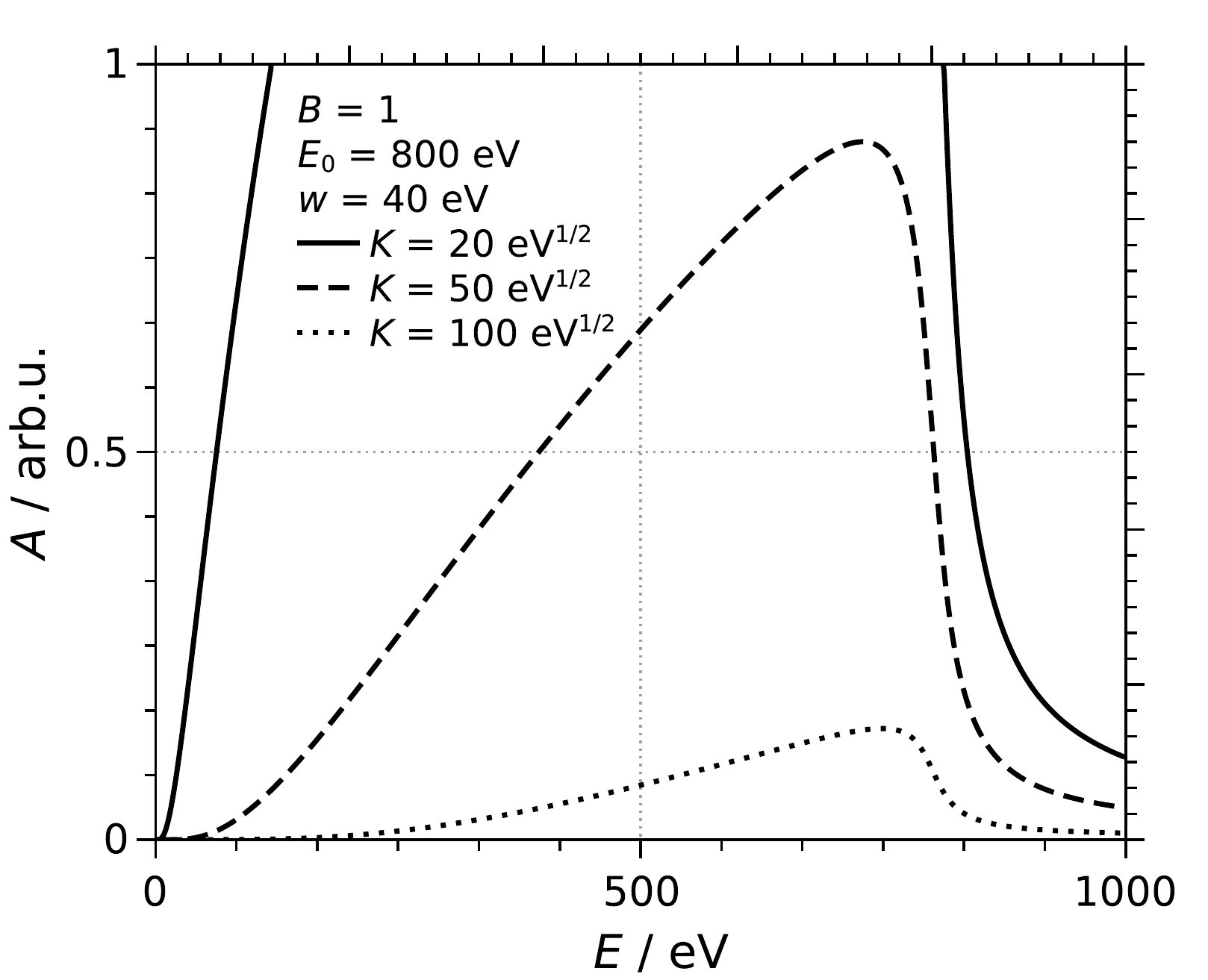}}
\caption{\label{fig:nelson} \doublespacing
Re-ionisation background calculated as product of equations \ref{eq:arctan} and \ref{eq:neutral} which is the original formula given by Nelson \cite{nel86}. Used parameter values are indicated in the figure.}
\end{figure}

In order to describe the shape of the re-ionisation continuum the original work of Nelson \cite{nel86} proposes to construct an analytic function combined from two different elements.
The first part is the arctangent function from the class of sigmoid functions.
Its purpose is to restrict the re-ionisation tail to the left side of the elastic scattering peak
\begin{equation}
\label{eq:arctan}
    f_{\rm arctan} = \pi - 2 \times \arctan(2 \times (E-E_{0})/w),
\end{equation}
where $E_{0}$ is the energy of the direct scattering peak and $w$ the full width at half maximum of the peak.
The decay toward smaller energies is modelled according to the probability that the ion is not neutralised in a binary collision \cite{bro07,dra05}
\begin{equation}
\label{eq:neutral}
    f_{\rm exp} = \exp(-v_{c} / v_{\perp}) = \exp(-K / {\rm sqrt}(E)),
\end{equation}
where $v_{c}$ is a critical velocity, $v_{\perp}$ the velocity of the ion perpendicular to the surface, $E$ the ion energy, and $K=v_{c} \times {\rm sqrt}(m) / ({\rm sqrt}(2) \times \cos(\theta))$ with $\theta$ being the angle of incidence of the ion with respect to the surface normal.
$v_{c}$, and therefore $K$, can in principle be determined from experiments where ion fractions are measured as function of $v_{\perp}$ \cite{pri08,pri09}.
Because data on $v_{c}$ is scarce and the measurements need considerable effort, $K$ is retained as a fitting parameter.
Combining both parts and multiplying with an amplitude factor $B$ yields the expression used in \cite{nel86}.
\begin{equation}
\label{eq:nelson}
    f_{\rm Nelson} = B \times \exp(-K/{\rm sqrt}(E)) \times (\pi - 2 \times \arctan(2 \times (E-E_{0})/w))
\end{equation}
Some representative curves are depicted in figure \ref{fig:nelson}.
The shown curves are plotted for varying values of $K$ and the same parameters $B$, $E_{0}$, and $w$.
It can be seen in the figure that the curves for different values of $K$ are extremely divergent in amplitude.
For example the curve with $K=\SI{100}{eV^{1/2}}$ shows rather small values, whereas the curve for $K=\SI{20}{eV^{1/2}}$ does not fit into the chosen plot window.
Although this behaviour could in principle be compensated by adjusting the amplitude factor $B$, it might cause problems during the parameter fitting procedure, because both parameters, $K$ and $B$, do strongly interfere with each other.

\begin{figure}
\resizebox{\figurewidth}{!}{\includegraphics{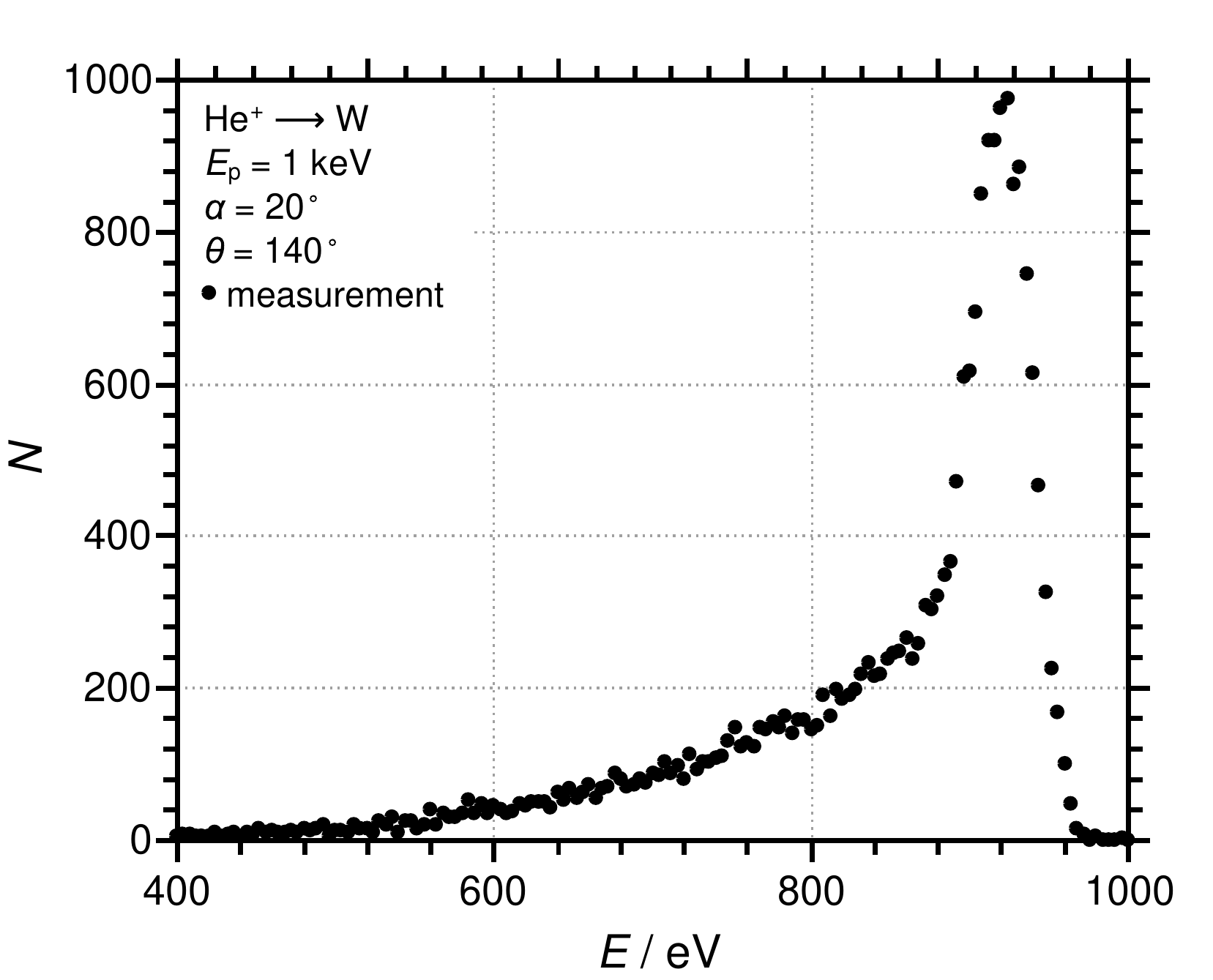}}
\caption{\label{fig:spec} \doublespacing
Measured low energy ion scattering spectrum of \SI{1}{keV} He$^{+}$ ions scattered from a poly-crystalline W surface under a scattering angle of \ang{140}.
The incident angle of the projectile ion beam is \ang{20} with respect to the surface normal.
The sample is kept at a temperature of $T=\SI{860}{K}$ during the measurement.}
\end{figure}

\begin{figure}
\resizebox{\figurewidth}{!}{\includegraphics{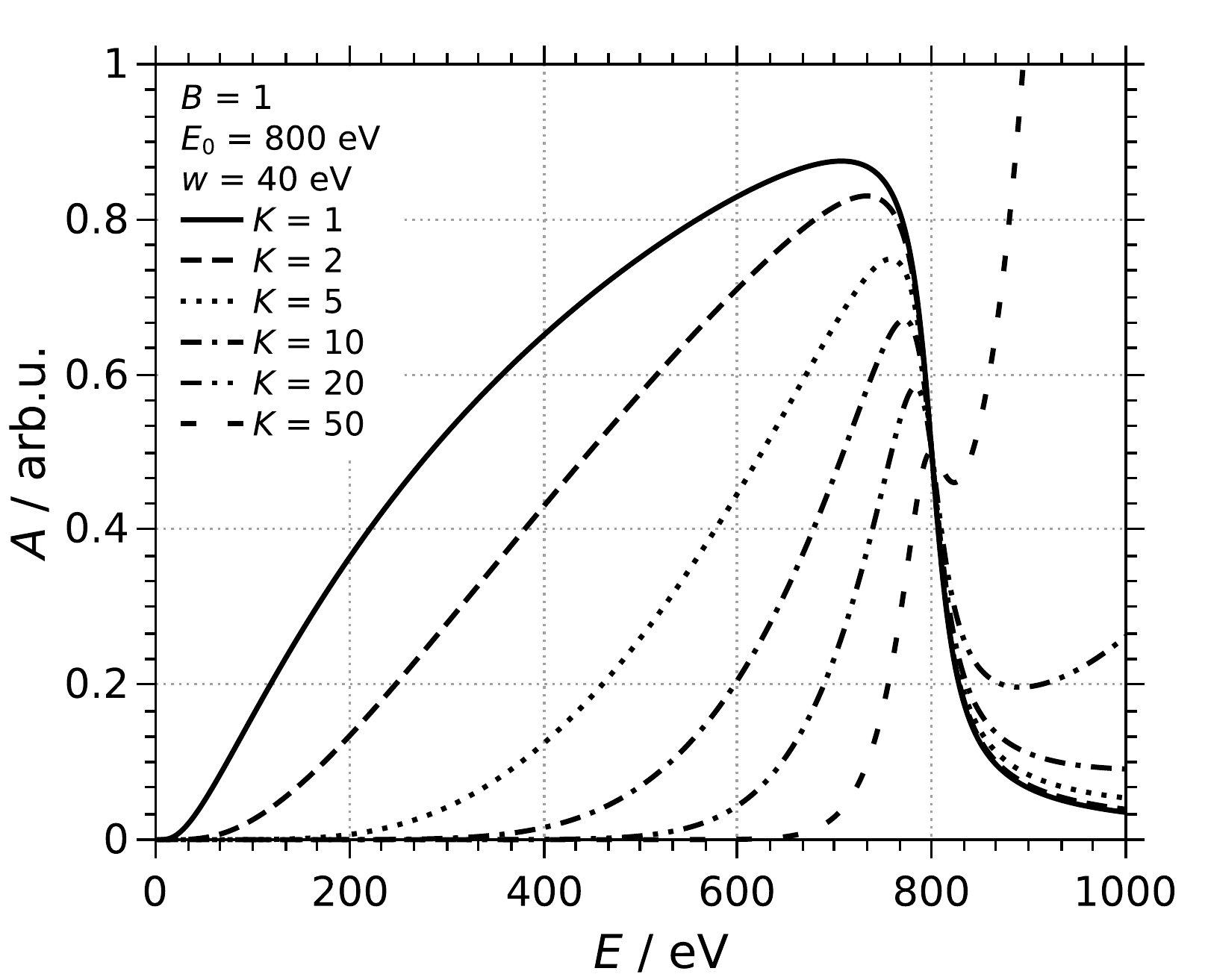}}
\caption{\label{fig:arctannorm} \doublespacing
Similar to figure \ref{fig:nelson} but including the normalisation shown in equation \ref{eq:arctannorm}.}
\end{figure}

\begin{figure}
\resizebox{\figurewidth}{!}{\includegraphics{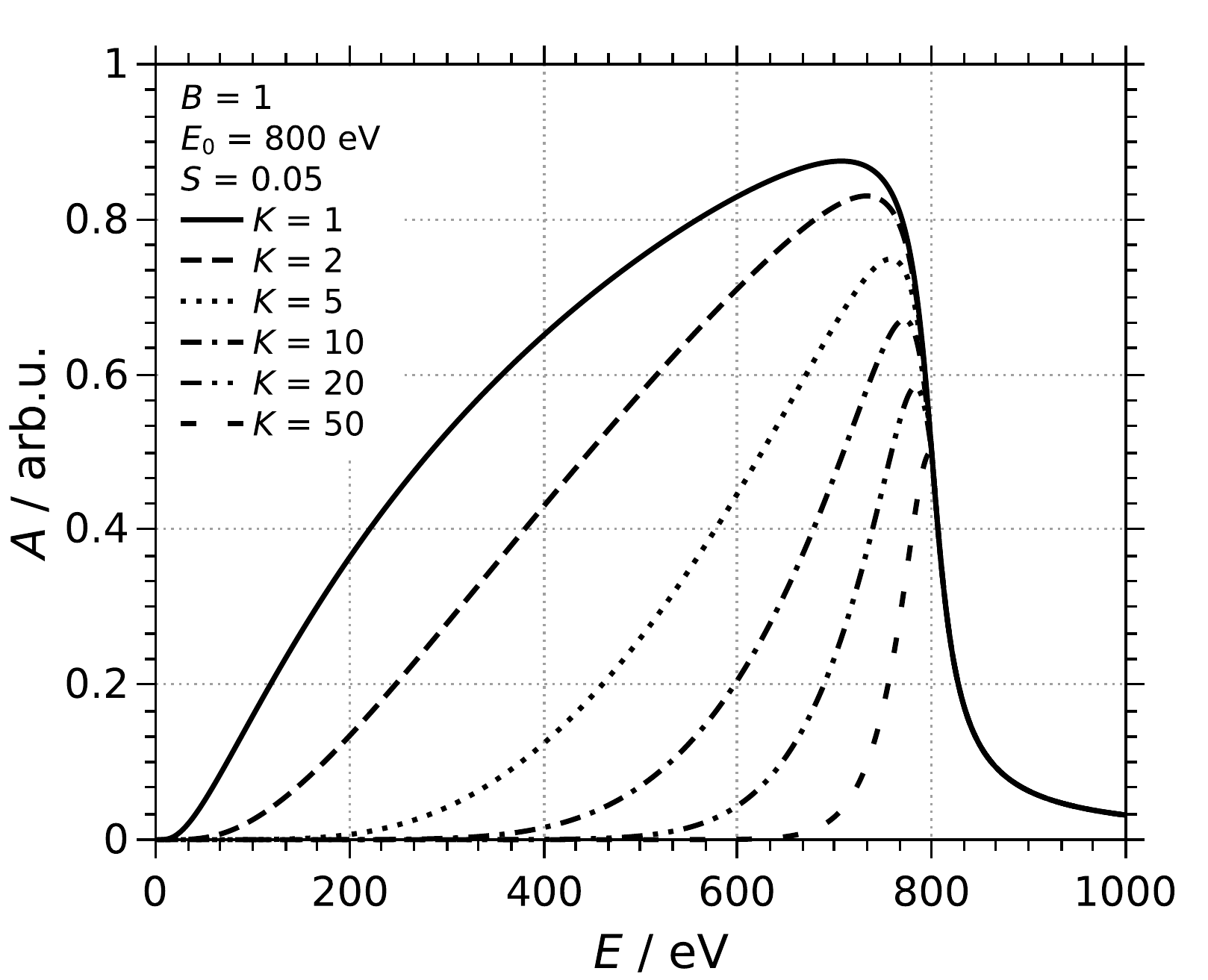}}
\caption{\label{fig:arctanconst} \doublespacing
Same as figure \ref{fig:arctannorm} but limiting the exponential function of the velocity dependence of neutralisation probability for $E>E_{0}$ according to equation \ref{eq:arctanconst}.}
\end{figure}

\begin{figure}
\resizebox{\figurewidth}{!}{\includegraphics{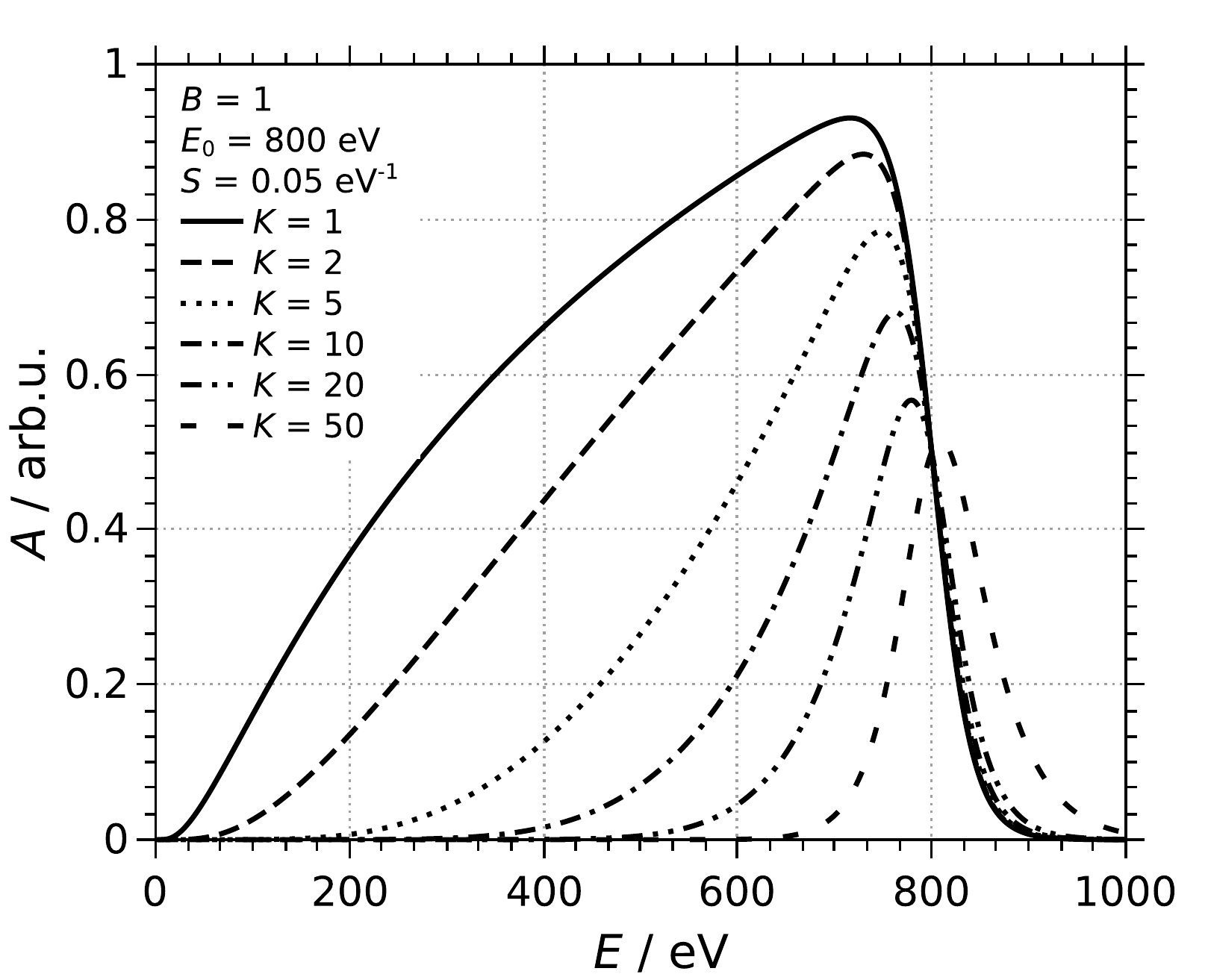}}
\caption{\label{fig:logistic} \doublespacing
Plots of the improved equation \ref{eq:logistic} for various values of $K$ showing the wide range of background shapes which can be fitted.}
\end{figure}

A feasible way to correct this unwanted situation is to find a proper normalisation of the background expression.
The aim is to modify the exponential (eq. \ref{eq:neutral}) and arctangent (eq. \ref{eq:arctan}) terms in a way to let the product of both define the shape of the background function whose amplitude is then given by $B$.
As a first step we map the arctangent term into the range \numrange{0}{1} by dividing by $2\pi$.
In a second step we modify the exponential term in equation \ref{eq:nelson}.
Again, in order to separate shape and values, the wide spread of curves with different $K$ values is removed by multiplying the term with $\exp{(K)}$, i.e. normalisation to 1 at $E=E_{0}$.
Furthermore, the independent variable, the particle energy $E$, is normalised by the energy $E_{0}$ of the corresponding direct scattering peak.
As result of these modifications one gets the following expression for the re-ionisation background:
\begin{equation}
\label{eq:arctannorm}
    f_{1} = B \times \exp{(K)} \times \exp{(-K \times {\rm sqrt}(E_{0}/E))} \times (\pi/2 - \arctan(2 \times (E-E_{0})/w)) / \pi.
\end{equation}
An array of curves according to this expression is plotted in figure \ref{fig:arctannorm}.
All parameters are the same as in figure \ref{fig:nelson}.
The plot reveals another artefact, which can be seen at energy values above the peak energy $E_{0}$.
For large values of $K$ (in excess of \numrange{10}{20}) the background expression diverges, i.e. the strong exponential increase can not be compensated and forced to decay toward \num{0} for large values of the energy.
This artefact can be rectified by a small amendment in equation \ref{eq:arctannorm}.
The strong rise of the exponential term at energies above the peak energy $E_{0}$ can be circumvented when the independent variable $E$ is limited to values $E \le E_{0}$.
With that modification the following expression for the re-ionisation background is obtained:
\begin{equation}
\label{eq:arctanconst}
    f_{2} = \left\{
	\begin{array}{ll}
	B \times \exp(K) \times \exp(-K \times {\rm sqrt}(E_{0}/E)) \times \dots \\
	    \dots (\pi/2 - \arctan(S \times (E-E_{0}))) / \pi      & E \le E_{0} \\
	B \times (\pi/2 - \arctan(S \times (E-E_{0}))) / \pi       & E > E_{0}
	\end{array} \right.
\end{equation}
Again, an array of curves is plotted in figure \ref{fig:arctanconst}.
The parameters are the same as in the preceding figure \ref{fig:arctannorm}.
The diverging feature is removed and all curves merge together into a unique curve for $E$ values above $E_{0}$.
The plot reveals another artefact which is caused by the arctangent term.
If one compares the shape of the background fit function with the actual shape of the measured spectrum in figure \ref{fig:spec}, one sees that the measured data decays rather quickly to zero at energy values above the peak energy.
In fact, this decay follows in general the shape of the Gaussian which is used to model the direct scattering peaks.
The spectrum fit using the background expression \ref{eq:arctanconst} is unlikely to yield an optimised match.
Replacing the arctangent function by the logistic function, another sigmoid type function, yields the final form of the re-ionisation background expression shown in the following equation:
\begin{equation}
\label{eq:logistic}
    f_{\rm bg} = B \times \exp{(K \times (1-{\rm sqrt}(E_{0}/E)))} \times (1-1/(1+\exp{(-S \times (E-E_{0}))})).
\end{equation}
The parameters $B$, $K$, and $E_{0}$ are the amplitude of the background, the exponential rise, and the energy of the direct scattering peak, as in the previous equations.
A fourth free parameter, $S$, is the slope of the logistic function at the position of the direct scattering peak.
This parameter is added in order to allow more flexibility and improves in general the fitting process of the data.
Figure \ref{fig:logistic} depicts the curves for the same parameters as in the previous plots, where the slope at the direct scattering peak position has been set to $S=1/w$ in order to give a close match.
The shape of the various curves shows a good agreement at energies below $E_{0}$, and show a much better convergence toward \num{0} at energies above $E_{0}$.
There is still a slight shift of the background function when the exponential parameter $K$ gets extreme values.
However, the experience with the use of the expression \ref{eq:logistic} shows that such large values are not encountered when fitting spectra as shown e.g. in figure \ref{fig:spec}.

\begin{figure}
\resizebox{\figurewidth}{!}{\includegraphics{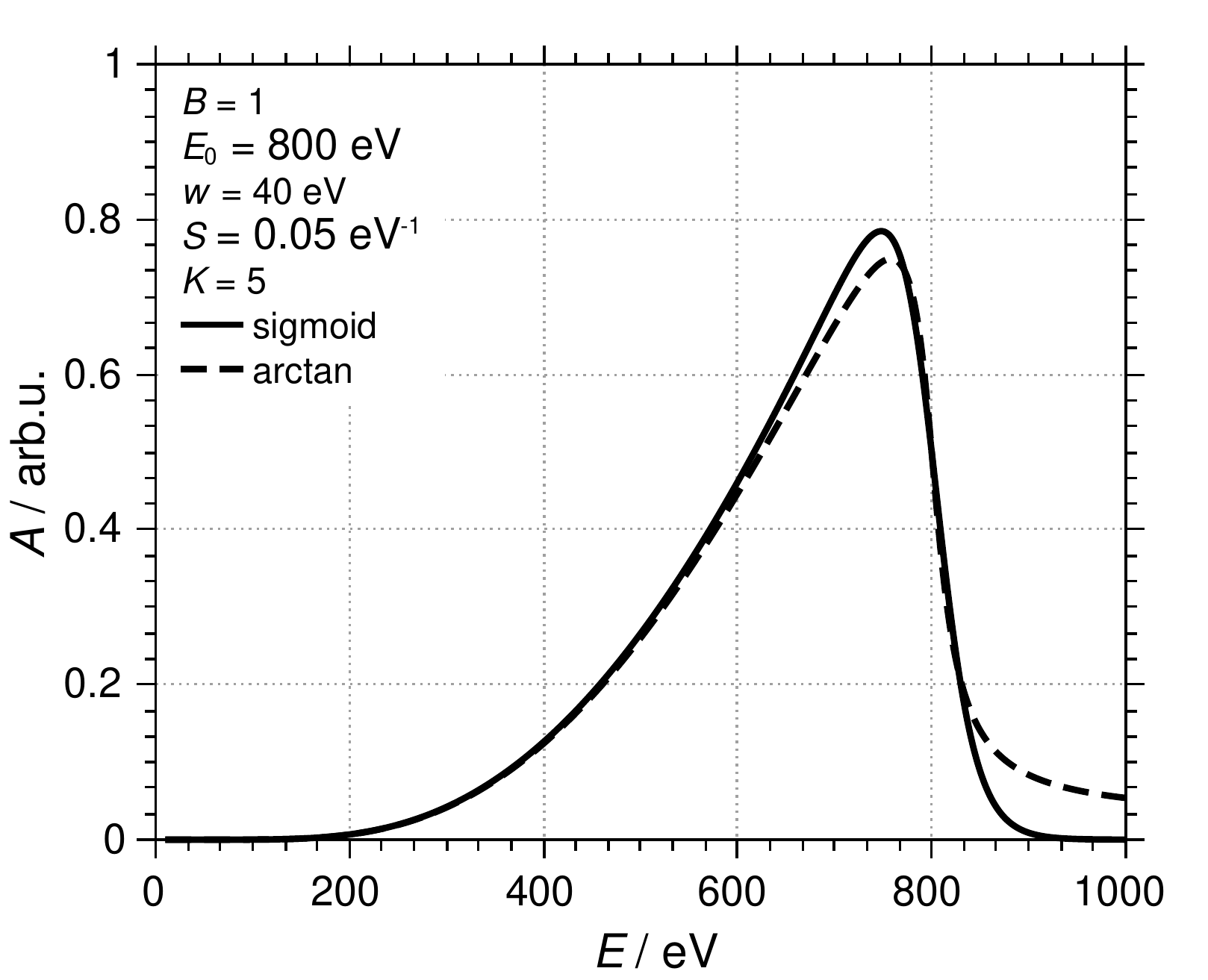}}
\caption{\label{fig:K5} \doublespacing
Comparison of background expression shape using the clipped arctangent (equation \ref{eq:arctanconst}) and the logistic (equation \ref{eq:logistic}) functions.}
\end{figure}

Two representative curves with $K=5$ from equations \ref{eq:arctanconst} and \ref{eq:logistic} are plotted in figure \ref{fig:K5}.
All parameters are the same, and $w$ and $S$ are matched by setting $S=1/w$.
There are only small deviations in the exponential increase at the left side, but it is clearly visible that the expression based on the logistic curve yields a much stronger decay to zero for $E \ge E_{0}$.

\section{Examples}
\begin{figure}
\resizebox{\figurewidth}{!}{\includegraphics{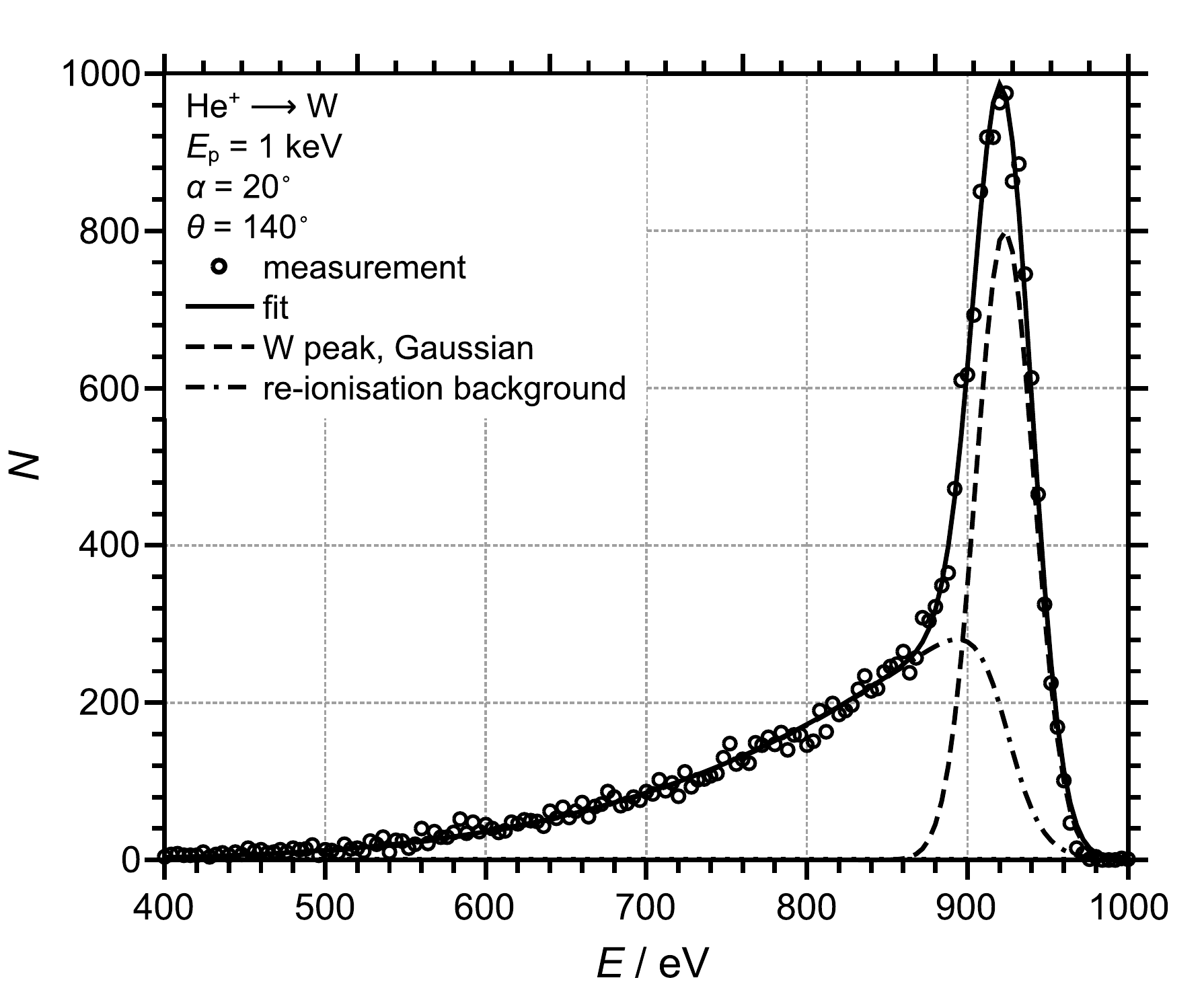}}
\caption{\label{fig:fit1} \doublespacing
Example of a combined fit (background and Gaussian peak) of the measurement shown in figure \ref{fig:spec}.}
\end{figure}

Figure \ref{fig:fit1} shows the same measurement (circles) as in figure \ref{fig:spec}, but this time a numerical fit of the data is added.
The direct scattering peak at $E_{0}=\SI{926}{eV}$ is fitted with a Gaussian curve $A \times \exp(-2.77259 \times ((E-E_{0})/w)^2)$\footnote[1]{In this notation of the Gaussian $A$ is the peak height and $w$ is the full width of half maximum.} (dashed line) and the long left tail resulting from re-ionisation events is fitted with the improved expression given in equation \ref{eq:logistic} (dash-dotted line).
Fitting is performed with a Levenberg-Marquardt non-linear regression algorithm using the \texttt{octave} \cite{octave} interactive programming environment.
The sum of both contributions (full line) does very well match the measured data.
Especially at energies larger than the direct scattering peak the decay toward \num{0} is fast enough and does not introduce any artefacts.

\begin{figure}
\resizebox{\figurewidth}{!}{\includegraphics{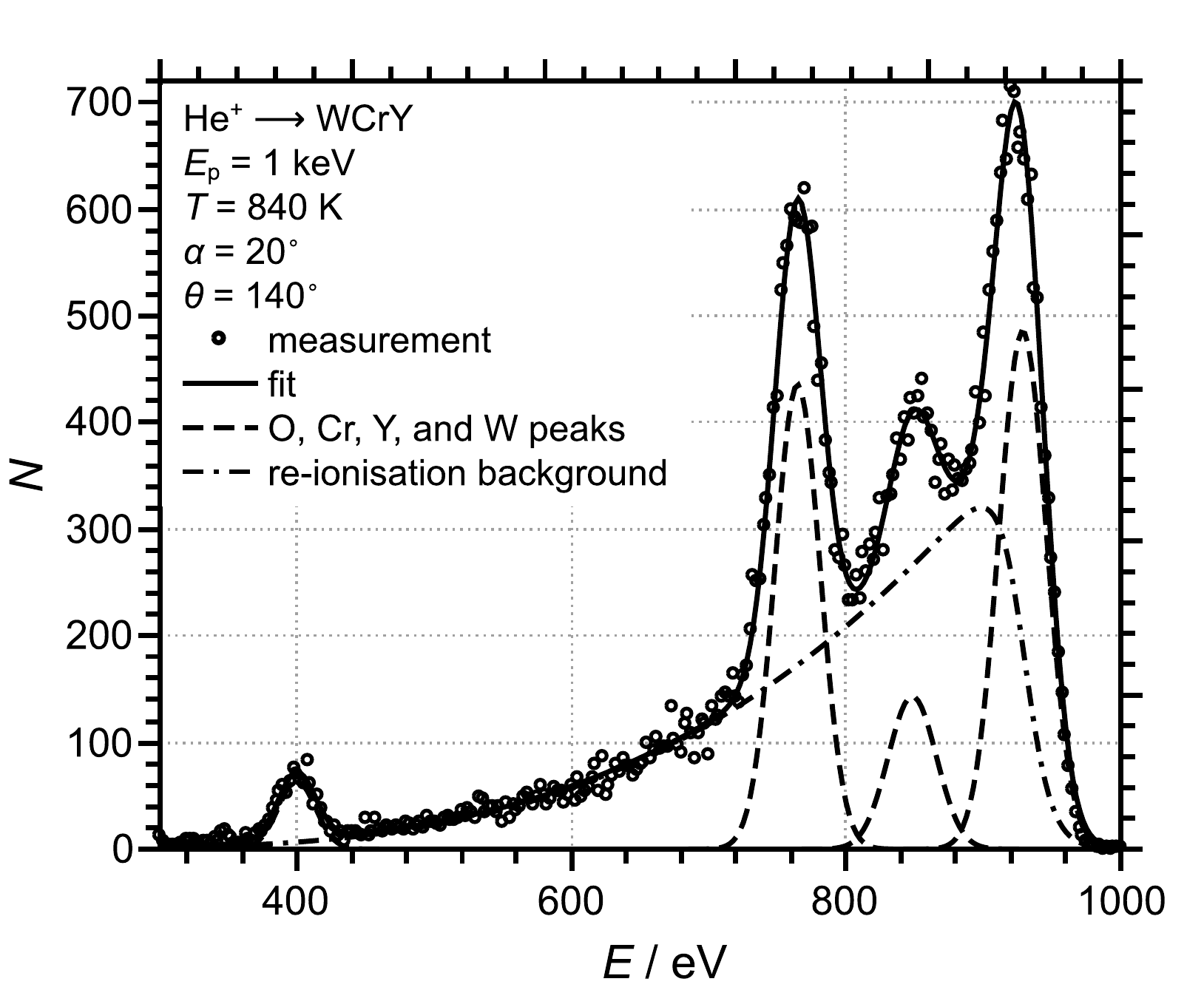}}
\caption{\label{fig:fit4} \doublespacing
A second example showing the fit of a more complex spectrum which is composed of four peaks and a re-ionisation continuum. The fitted curve (full line) is the sum of four Gaussians and the improved expression for the re-ionisation background.}
\end{figure}

A second, more complex spectrum is shown in figure \ref{fig:fit4}.
The plot shows the counts (circles) of He$^{+}$ ions scattered from the surface of a W--11.4Cr--0.6Y alloy (numbers are weight\%) where the temperature driven segregation of Cr to the surface has been studied in detail \cite{kos20}
\footnote[2]{Note that the measurements shown in \cite{kos20} have been analysed with the background expression \ref{eq:arctanconst} still using the arctangent sigmoid function.}.
The spectrum shows four peaks which are attributed to O, Cr, Y, and W on the surface of the sample.
In addition, the W peak exhibits the re-ionisation tail at the left side which lifts up the Cr and Y peaks.
The measured spectrum is fitted with the sum of four Gaussians (dashed lines) and the re-ionisation expression from equation \ref{eq:logistic} (dash-dotted line).
For completeness a constant offset (which gives a very small correction) is fitted, too.
Table \ref{tab:fit} lists the fitting parameters and the determined values together with errors derived in the usual way from the diagonal elements of the covariance matrix.
Again, the sum of all considered contributions (full line) gives a very good representation of the measured data.
The re-ionisation threshold for Y is only \mbox{\SI{300}{eV}} \cite{sou87} and the threshold for Cr is below \mbox{\SI{1000}{eV}} \cite{tho86}, so both should strictly be included in the fitting procedure.
Adding re-ionisation tails for Cr and Y does not result in good convergence because the overlapping exponential functions generate an ill-posed problem.
However, comparing figures \ref{fig:fit1} and \ref{fig:fit4} reveals that the re-ionisation background in the latter example extends to lower scattering energies (indicated by the smaller $K$ value in table \ref{tab:fit}), obviously fitting the re-ionisation from all constituents.

\section{Discussion and Conclusion}
The numerical fitting of LEIS spectra is a necessary prerequisite when surface densities of the various constituents of the sample under investigation should be obtained.
For that purpose the amplitudes (areas) of the direct scattering peaks need to be determined.
Contributions resulting from scattering events in deeper layers have to be subtracted which requires a correct estimate.
To our knowledge a quantitative description of the re-ionisation background is not available in the literature.
The semi-empirical approach brought forward by Nelson \cite{nel86} does at least include the functional dependence of neutralisation processes on the particle energy.
Through application of this formula a couple of numerical artefacts are noticed.
These artefacts could be removed by various modifications deriving the improved expression for the re-ionisation background.
There could be other numerical procedures which might work for the purpose of spectrum fitting.
For example in \cite{avv19} the authors use sloped error functions to quantify the background.
Although this approach seems to be feasible in many situations, it remains doubtful if a complex spectrum like the one shown in figure \ref{fig:fit4} could be equally well matched because it is questionable if this method can handle the subtraction of convex shaped background functions as encountered in figure \ref{fig:fit1}.

\newpage
\section*{Acknowledgements}
This work has been carried out within the framework of the EUROfusion Consortium and has received funding from the Euratom research and training programme 2014-2018 and 2019-2020 under grant agreement No 633053. The views and opinions expressed herein do not necessarily reflect those of the European Commission.

\section*{Data availability statement}
The shown experimental data, the results of the fitting procedure, and the \texttt{octave} program for spectrum fitting using the new expression for the re-ionisation background can be downloaded from the J\"ulich DATA repository \url{https://doi.org/10.26165/JUELICH-DATA/2GRSPN}.

\section*{ORCID IDs}
H. R. Koslowski \url{https://orcid.org/0000-0002-1571-6269}

Ch. Linsmeier \url{https://orcid.org/0000-0003-0404-7191}

\newpage
\begin{table}[h]
 \begin{tabular}[h]{l l r r l}
  \hline{}
   target & parameter &   value &        error & unit \\
  \hline{}
   W      & $A_{W}$  &    801.0 &   $\pm 15.6$ &  \\
          & $E_{W}$  &    923.3 &    $\pm 1.3$ & eV \\
          & $w_{W}$  &    42.51 &   $\pm 0.66$ & eV \\
          & $B_{W}$  &    348.1 &    $\pm 8.0$ &  \\
          & $K_{W}$  &     9.53 &   $\pm 0.36$ &  \\
          & $S_{W}$  &   0.0947 & $\pm 0.0582$ & eV$^{-1}$ \\
          & offset   &     0.93 &   $\pm 2.35$ &  \\
  \hline{}
   WCrY   & $A_{O}$  &     63.7 &    $\pm 5.5$ &  \\
          & $E_{O}$  &    399.8 &    $\pm 1.3$ & eV \\
          & $w_{O}$  &    32.07 &   $\pm 3.35$ & eV \\
          & $A_{Cr}$ &    435.1 &    $\pm 5.9$ &  \\
          & $E_{Cr}$ &    764.5 &    $\pm 0.2$ & eV \\
          & $w_{Cr}$ &    38.85 &   $\pm 0.66$ & eV \\
          & $A_{Y}$  &    143.2 &    $\pm 9.1$ &  \\
          & $E_{Y}$  &    848.1 &    $\pm 0.8$ & eV \\
          & $w_{Y}$  &    41.29 &   $\pm 3.35$ & eV \\
          & $A_{W}$  &    485.1 &   $\pm 24.9$ &  \\
          & $E_{W}$  &    928.5 &    $\pm 2.1$ & eV \\
          & $w_{W}$  &    41.00 &   $\pm 0.85$ & eV \\
          & $B_{W}$  &    380.4 &   $\pm 18.2$ &  \\
          & $K_{W}$  &     7.75 &   $\pm 0.33$ &  \\
          & $S_{W}$  &   0.1033 & $\pm 0.0430$ & eV$^{-1}$ \\
          & offset   &     1.79 &   $\pm 2.22$ &  \\
  \hline
 \end{tabular}
 \caption{\label{tab:fit} \doublespacing{}
 Fitting results for the spectra shown in figures \ref{fig:fit1} (top) and \ref{fig:fit4} (bottom). $A$ are the amplitudes, $E$ the peak energies, and $w$ the full widths of half maximum of the four Gaussian peaks.
 The respective elements are indicated as subscripts.
 $B$, $K$, and $S$ are the free fitting parameters of the re-ionisation background according to equation \ref{eq:logistic}.}
\end{table}

\newpage

\newpage

\newpage

\newpage

\end{document}